\documentclass{iau}
\usepackage{graphicx}
\usepackage{aas_macros}
\usepackage{amsmath}
\usepackage{graphicx}
\usepackage{wrapfig}
\usepackage{orcidlink}
\lefttitle{N. Saha \textit{et al.}}
\righttitle{The effect of triaxiality on the dynamics of triple supermassive
black holes in a cosmological context}
\volno{398}
\jnlPage{1}{7}
\jnlDoiYr{2025}
\doival{10.1017/xxxxx}

\aopheadtitle{Proceedings IAU Symposium}
\editors{Hyung Mok Lee, Rainer Spurzem \& Jongsuk Hong, eds.}

\title{The effect of triaxiality on the dynamics of triple supermassive black holes in a cosmological context}

\author{Navonil Saha$^1\orcidlink{0000-0002-7304-6865}$, Peter Berczik$^{2,3}\orcidlink{0000-0003-4176-152X}$, Andreas Just$^1\orcidlink{0000-0002-5144-9233}$ and Margarita Sobolenko$^{2,3}\orcidlink{0000-0003-0553-7301}$}

\affiliation{{$^1$Zentrum f\"ur Astronomie der Universit\"at Heidelberg, Astronomisches Rechen-Institut,\\M\"{o}nchhofstr. 12-14, 69120, Heidelberg, Germany\\ email: \email{navonil.saha@uni-heidelberg.de}}\\
{$^2$Nicolaus Copernicus Astronomical Centre, Polish Academy of Sciences,\\ul. Bartycka 18, 00-716 Warsaw, Poland}\\
{$^3$Main Astronomical Observatory, National Academy of Sciences of Ukraine, \\27 Akademika Zabolotnoho St, 03143 Kyiv, Ukraine}}

\begin{document}

\begin{abstract}
The hierarchical nature of galaxy formation in the $\Lambda$CDM framework often leads to multiple supermassive black holes (SMBHs) in the galactic nuclei. The timescale over which galaxies merge, plays a crucial role in shaping the dynamical evolution and the merger dynamics of their central SMBHs. While binary SMBH evolution is well studied, the long-term dynamics of triple SMBH systems, particularly in non-spherical potentials, remain less understood. We investigate the role of triaxiality in the evolution and dynamics of triple SMBHs with initial conditions drawn from the ROMULUS25 cosmological simulation, using high-resolution gravitodynamical $\textit{N-body}$ simulations. We explore different orbital configurations and host shapes, tracking the evolution from galactic inspiral to hard binary formation at sub-parsec scales. In all cases, the two most massive SMBHs form a rapidly hardening binary that coalesces within a fraction of a Hubble time, while the third forms a stable hierarchical triple system with the heavier binary, or remains on a wide orbit. 

\end{abstract}


\maketitle
\section{Introduction}
During their lifetimes, most massive galaxies undergo various major merger events, either with other individual galaxies or with groups of galaxies. This has been the basis of the hierarchical cosmic structure formation in the $\Lambda$CDM cosmology \citep{1970ApJ...162..815P, 1978MNRAS.183..341W, 1991ApJ...379...52W}. Galaxies gain mass through hierarchical mergers, and most elliptical and spherical galaxies host a central supermassive black hole (hereafter, SMBH). The emperical scaling relations suggest that the host galaxies and their central SMBHs grow together and suggest a co-evolutionary connection between them like the $M_{BH}$–$\sigma$ relation which links the SMBH mass $M_{BH}$ to the stellar velocity dispersion $\sigma$ of the galactic bulge \citep{2000ApJ...539L...9F, 2000ApJ...539L..13G}. When the galaxy grows, $\sigma$ increases and hence subsequently the $M_{BH}$. The SMBHs undergoes stellar dynamical evolution through mergers of galaxies that leads to the formation of a SMBH binary and its eventual coalescence with the probable emission of gavitational waves detectable by current and future observatories \citep{2017arXiv170200786A}.

The dynamical evolution of the SMBH is commonly divided into three distinct stages, governed by different physical mechanisms \citep{1980Natur.287..307B}. First, in the post merger phase of the galaxy mergers, the SMBHs sink to the center of the galaxy through dynamical friction with other stellar bodies and particles \citep{1943ApJ....97..255C}. The gravitational drag decelerates the SMBHs, causing them to lose orbital energy and angular momentum. This process operates efficiently on kiloparsec to parsec scales and continues until the two SMBHs form a bound binary. Once the SMBHs form a bound binary, further orbital decay is dominated by interactions with the surrounding stellar environment and this is known as the stellar hardening phase. The pair thus forms a hard binary and continue ejecting stellar particles through three-body encounters (as gravitational slingshot), carrying away energy and angular momentum \citep{1996NewA....1...35Q, 2010ApJ...719..851S}. This process gradually reduces the binary separation, a stage often referred to as “hardening” and causes the binary to shrink. Finally, at sub-parsec scales, the emission of gravitational waves dominates, rapidly driving the binary to coalescence \citep{1963PhRv..131..435P, 1964PhRv..136.1224P}.

The timescale from the merger of galaxies to the final coalescence of their central SMBHs can vary significantly, depending on the remnant’s matter distribution and the binary’s orbital parameters, ranging from tens of Myr to nearly a Gyr or longer \citep{2012ApJ...749..147K, 2016ApJ...828...73K, 2014SSRv..183..189C}. During this evolution, it is possible that a third galaxy joins this galaxy, which already contains a SMBH binary and create a triple SMBH system, or even three (or more) galaxies directly merge simultaneously. However, the long term dynamics of triple SMBH systems still remains poorly understood. In previous work by \cite{2023A&A...678A..11K}, triple SMBH systems, where two subsequent merger takes places within a Gyr, were extracted from the merger tress of SMBHs and it was found that the triple interactions plays a crucial role in the outcome of the dynamical evolution of the SMBHs. In these simulations spherical galaxy systems were used with the initial conditions taken from the ROMULUS25 cosmological simulation \citep{2017MNRAS.470.1121T}, and it turned out that the two most massive SMBHs merge first.  In this work, we extend these models on triple SMBHs by investigating the role of triaxiality in shaping the dynamical evolution of three SMBHs systems to find the common dynamical evolution pattern of the triple system and predict the typical coalescence time scale using high resolution gravitodynamical N-body simulations.

\section{Methods and initial conditions}
We use the spherical galaxy models of \cite{2023A&A...678A..11K} in order to create different models of 3-galaxy triaxial initial conditions with dark matter, gas and stars taking the 3 galaxy radial galactic Dehnen profile parameters along with the SMBH initial conditions from Table 2 and Table 3 of the paper. The masses of the SMBHs are $88.4\times10^7M_\odot , 13.3\times10^7M_\odot\ \text{and}\ 3.6\times10^7M_\odot$ respectively for BH1, BH2 and BH3. All the three models A, B and C in this work (corresponding to model A1, A2 and A3 in \cite{2023A&A...678A..11K}) differ in the initial galactic density profiles of the stellar components. We then add the triaxial shape taken from \cite{2016MNRAS.462..663B} to these galaxies using AGAMA galaxy modeling software \citep{2019MNRAS.482.1525V}. Here the potential of the SMBHs were not taken into account in order to create in a few dynamical time scales of the nucleus in the simulation, a central cusp of the central potential of the SMBH.

We use different high resolution N-body codes for our simulations. For the initial 3-body interaction phase we start with the BONSAI2 Tree code \citep{2012JCoPh.231.2825B} with $\sim 10$ million particles for each of the galaxy for sufficient interaction of the particles during the early evolution of the global galaxy merger phase and the decay of the SMBHs to the center of the merger remnant due to dynamical friction. Once the binary starts to harden and the separation of one black hole pair fall below $\sim$100 pc, the risk of close pericenter passages not being properly resolved due to the lack of uniform timestep starts to grow and we get a time resolution problem. Therefore, at this point we shift to the $\varphi$-GPU Direct N-body code \citep{2011hpc..conf....8B} with an individual black timestep scheme for the hardening phase of SMBHs and correctly resolve the 3-body dynamics close to SMBHs until we get a bound system with a hardening binary. In this step we also reduce the total number of particles from $\sim30\times10^6$ to $\sim4.8\times10^6$ particles close to the inner 10 kpc of the SMBHs. We also include Post-Newtonian (PN) approximation up to 2.5 PN order when the Black holes reaches separation of 1000$R_s$ (schwarzschild radius) and move towards the relativistic regime. 
\section{Results}
\begin{figure}[h]\vspace{-0.42cm}
    \centering
        \includegraphics [width=\linewidth]{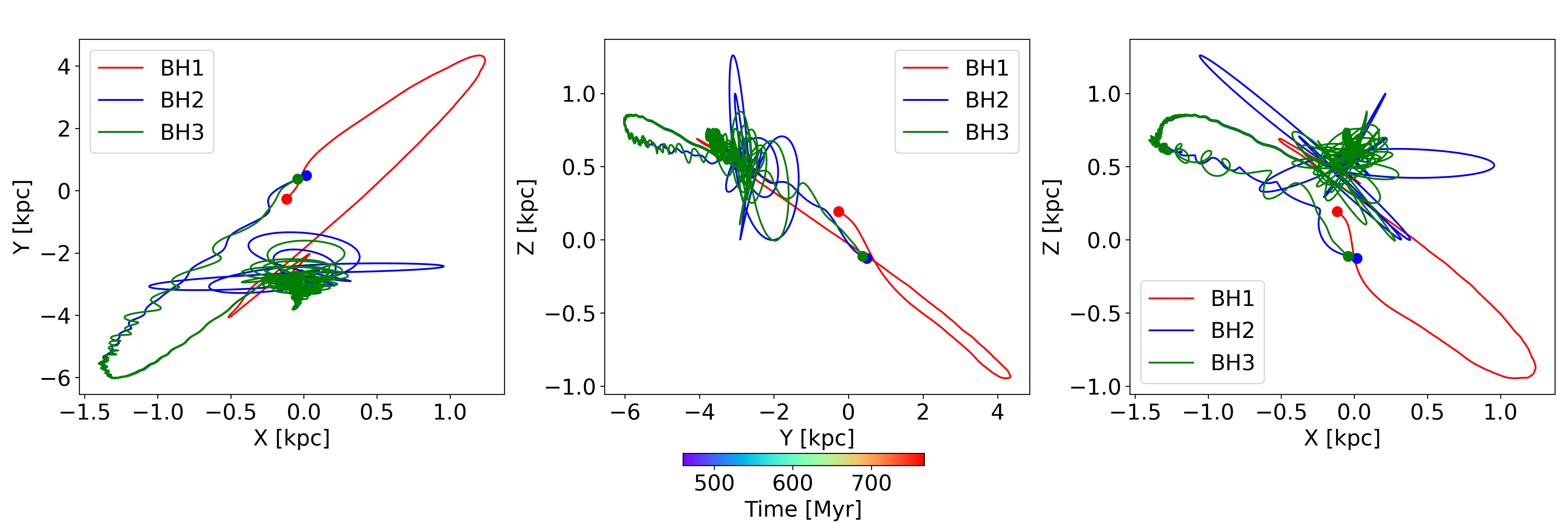}
    \vspace{-0.77cm}
    \caption{Orbital evolution of the three SMBHs where the three BH1,BH2 \& BH3 are color-coded as red, blue and green respectively showing the continuous exchange of the SMBHs due to three body encounters.}
    \label{fig:orbital evolution}
\end{figure}

The orbital evolution of the triple SMBH system showed that they form a binary which have orbits around the most massive BHs. From time to time there are three body encounters of the SMBHs which switch the members of the binary for all our models (see Fig. \ref{fig:orbital evolution}).
\vspace{-0.75cm}
\begin{figure}[h]
    \centering
    \includegraphics[width=\linewidth]{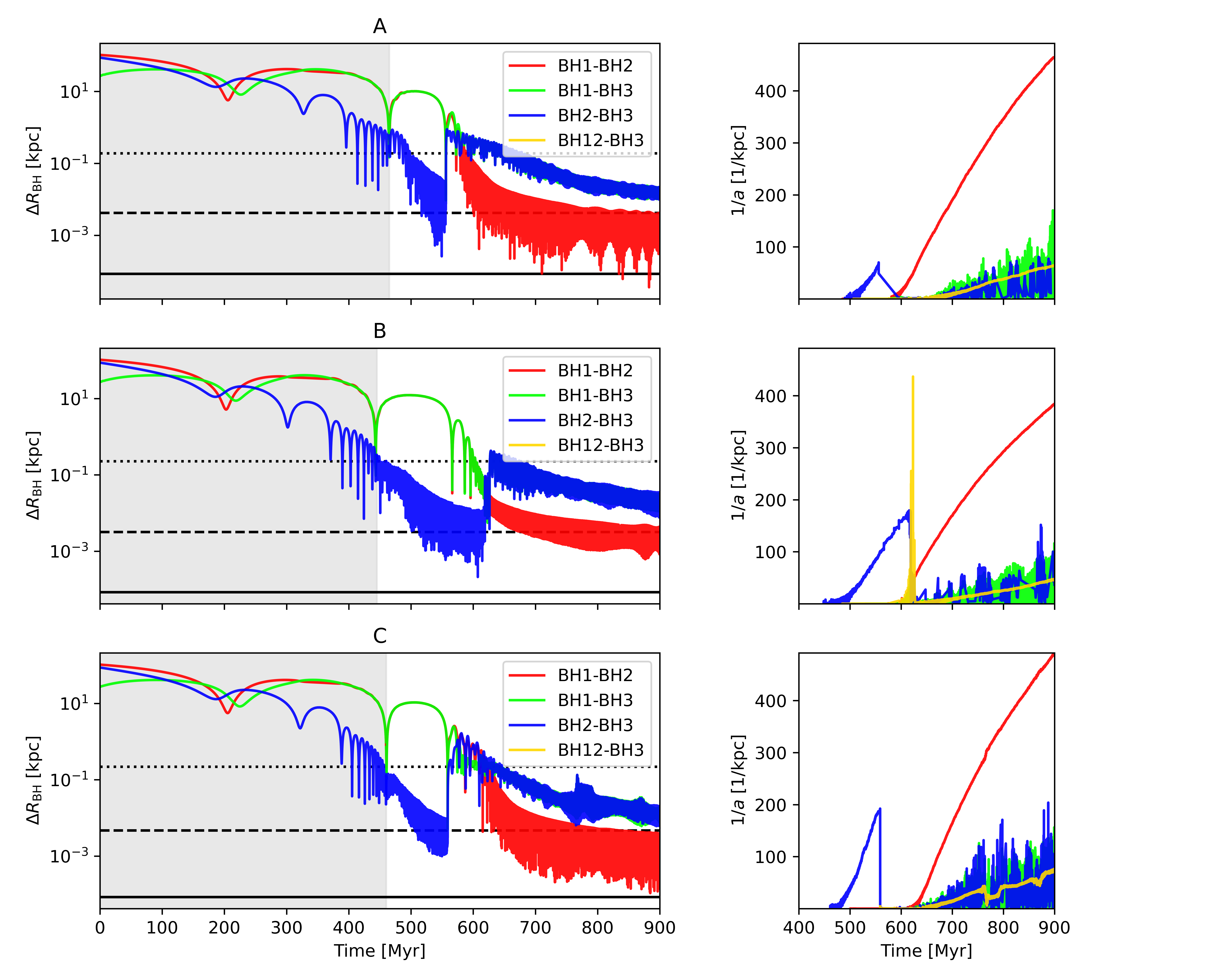}
    \vspace{-0.5cm}
    \caption{Black hole orbital parameters across the different simulations. The color coding refers to different SMBH pairs. The red lines refer to the orbital parameters around the inner BH1-BH2 binary (two most massive BHs). Background colors mark the codes used during the particular simulation period, light gray for Bonsai2, white for $\varphi$-GPU. The left panels show the separation of different SMBH pairs and the right panels show the evolution of the inverse semi-major axis. The black dotted, dashed, and solid lines, mark the influence radius, hardening radius, and the factor 1000 of the Schwarzschild radius ($R_s$) of BH1 respectively.} %\vspace{-0.15cm}
    %Figure taken from Saha et. al., 2025 (in prep.).}
    \label{fig:Rbh_and_1_a_modelA}
\end{figure}\\ 
For all the three models, initially the BH2 and BH3 approach each other very closely, which then later during the final galactic collision weakens the BH2 slingshot. Hence, the BH2 immediately returns to the galactic center and we get a BH2-BH3 system (blue line) sharing a common halo and orbiting each other,  preventing the formation of the BH1–BH3 system (green line). Now as the separation starts to fall below sub-pc scale we switched to the direct code (marked in white) between 400-500 Myr. Between 500-700 Myr there is a collision of the BH2-BH3 with the first galaxy and we get an exchange and finally a BH1–BH2 binary (red line) forms, as BH3 is sent off to orbit around the galactic center at 100–500 pc. Now, while the BH1–BH2 binary continues to harden, BH3 slowly descends, its orbit  circularizes and a stable hierarchical triple system is formed in all the models, where BH3 becomes bound to the inner BH1–BH2 hard binary (the two most massive SMBHs) as shown in Fig. \ref{fig:Rbh_and_1_a_modelA}.

We take the orbital averaged hardening rate by GW emission from the empirical \cite{1963PhRv..131..435P} formula assuming a constant eccentricity and add the stellar hardening estimated from the simulation and calculate the merging time of the binary. We get a stellar hardening rate between 1-2 kpc$^{-1}$Myr$^{-1}$ for the hard binary and merging time between $\sim$3-13 Gyr which is well within the Hubble time. The real merging rate of the models is different because of the different hardening rate of the binary system, which actually depends on the details of the orbit. Hence, we conclude that triaxiality does not show any strong impact on the interaction and the final outcome of the triple system as compared to SMBHs present in spherical galaxies.

\begin{acknowledgements}
The authors gratefully acknowledge the Gauss Center for Supercomputing eV (www.gauss-centre.eu) for funding this project by providing computing time through the John von Neumann Institute for Computing (NIC) on the GCS Supercomputer JUWELS at Jülich Supercomputing Center (JSC). The authors acknowledge support by the High Performance and Cloud Computing Group at the Zentrum für Datenverarbeitung of the University of Tübingen, the state of Baden-Württemberg through bwHPC and the German Research Foundation (DFG) through grant no INST 37/1159-1 FUGG, and the data storage service SDS@hd supported by the Ministry of Science, Research and the Arts Baden-Württemberg (MWK) and the German Research Foundation (DFG) through grant INST 35/1503-1 FUGG. NS acknowledges the support of doctoral scholarship from the German Academic Exchange Service (DAAD) through grant no. 91864951. PB and MS thanks the support from the special program of the Polish Academy of Sciences and the U.S. National Academy of Sciences under the Long-term program to support Ukrainian research teams grant No. PAN.BFB.S.BWZ.329.022.2023. 
\end{acknowledgements}
\vspace{-0.46cm}
\bibliographystyle{iaulike}
\bibliography{Sample}

%\begin{thebibliography}{}
%\bibitem[Bouvier(2013)]{2013EAS....62..143B} Bouvier, J.\ 2013, EAS Publications Series, 143
%\bibitem[Collier Cameron(1999)]{1999ASPC..158..146C} Collier Cameron, A.\ 1999, Solar and Stellar Activity: Similarities and Differences, 146
%\bibitem[Donati~{\it et. al}(1992)]{1992A&A...265..682D} Donati, J.-F., Brown, S.~F., Semel, M., {\it et. al}\ 1992, {\it A\&A}, 265, 682
%\end{thebibliography}

\end{document}